\documentclass{article}
\pdfoutput=1
\usepackage{jcappub}
\usepackage{graphicx}
\usepackage[latin1]{inputenc}
\usepackage{amsmath}
\usepackage{amssymb}
\usepackage{a4wide}
\usepackage{url}
\def\simlt{\ \raise -2.truept\hbox{\rlap{\hbox{$\sim$}}\raise5.truept   %
\hbox{$<$}\ }}
\def\simgt{\ \raise -2.truept\hbox{\rlap{\hbox{$\sim$}}\raise5.truept   %
\hbox{$>$}\ }}                                                          %

\def\be{\begin{equation}}
\def\ee{\end{equation}}
\def\newline{\hfil\break}

\def\la{\mathrel{\hbox{\rlap{\hbox{\lower4pt\hbox{$\sim$}}}\hbox{$<$}}}}
\def\ga{\mathrel{\hbox{\rlap{\hbox{\lower4pt\hbox{$\sim$}}}\hbox{$>$}}}}

\newcommand{\pd}[3]{\frac{\partial^{#3} #1}{\partial {#2}^{#3}}} 
\newcommand{\td}[3]{\frac{d^{#3} #1}{d {#2}^{#3}}} 
\renewcommand{\v}[1]{\ensuremath{\mathbf{#1}}} 
\newcommand{\gv}[1]{\ensuremath{\mbox{\boldmath$ #1 $}}}
\renewcommand{\bar}[1]{\ensuremath{\overline{#1}}}

\title{A Multi-frequency analysis of possible Dark Matter Contributions to M31 Gamma-Ray Emissions.}
\author[a,1]{G. Beck\note{Corresponding author.}}
\author[a]{S. Colafrancesco}

\affiliation[a]{School of Physics, University of the Witwatersrand, Private Bag 3, WITS-2050, Johannesburg, South Africa}

\emailAdd{geoffrey.beck@wits.ac.za}
\emailAdd{sergio.colafrancesco@wits.ac.za}

\abstract{We examine the possibility of a dark matter (DM) contribution to the recently observed gamma-ray spectrum seen in the M31 galaxy. In particular, we apply limits on Weakly Interacting Massive Particle DM annihilation cross-sections derived from the Coma galaxy cluster and the Reticulum II dwarf galaxy to determine the maximal flux contribution by DM annihilation to both the M31 gamma-ray spectrum and that of the Milky-Way Galactic Centre. We limit the energy range between $1$ and $12$ GeV in M31 and Galactic Centre spectra due to the limited range of former's data, as well as to encompass the high-energy gamma-ray excess observed in the latter target. In so doing, we will make use of Fermi-LAT data for all mentioned targets, as well as diffuse radio data for the Coma cluster. The multi-target strategy using both Coma and Reticulum II to derive cross-section limits, as well as multi-frequency data, ensures that our results are robust against the various uncertainties inherent in modelling of indirect DM emissions.

Our results indicate that, when a Navarro-Frenk-White (or shallower) radial density profile is assumed, severe constraints can be imposed upon the fraction of the M31 and Galactic Centre spectra that can be accounted for by DM, with the best limits arising from cross-section constraints from Coma radio data and Reticulum II gamma-ray limits. These particular limits force all the studied annihilation channels to contribute $1\%$ or less to the total integrated gamma-ray flux within both M31 and Galactic Centre targets. In contrast, considerably more, $10-100$\%, of the flux can be attributed to DM when a contracted Navarro-Frenk-White profile is assumed. This demonstrates how sensitive DM contributions to gamma-ray emissions are to the possibility of cored profiles in galaxies. The only channel consistently excluded for all targets and profiles (except for $\sim 10$ GeV WIMPs) is the direct annihilation into photons.

Finally, we discuss the ramifications of evidence in favour of cored halo density profiles for DM explanations of galactic gamma-ray emission.}

\begin{document}
\maketitle

\section{Introduction}

The Galactic Centre of the Milky-Way (GC) has been of great interest in attempts to indirectly detect Dark Matter (DM). This is due to the existence of a spherically symmetric excess in both high-energy and very-high-energy gamma-rays that has been observed within its environs~\cite{fermiexcess,hessexcess} by experiments like the Fermi Large Area Telescope~\cite{fermi-docs} (Fermi-LAT) and the High Energy Stereoscopic System~\cite{hess-details} (HESS). The advent of this observational excess gave rise to many attempts to explicate it in terms of emission resulting from the annihilation of DM in the form of Weakly Interacting Massive Particles (WIMPs)~\cite{dmgc1,dmgc2,dmgc3,dmgc4,dmgc5,dmgc6,dmgc7,dmgc8,dmgc9}. However, alternative explanations, such as unresolved populations of milli-second pulsars, have been extensively discussed in the literature~\cite{oleary2015,bartels2015b,lee2015,brandt2015,calore2015} and have received recent reinforcement~\cite{ploeg2017}, as it has been shown that the same luminosity function could account for both known and unresolved populations of pulsars. In addition to this, the region of the WIMP parameter space favoured by analysis of astrophysical uncertainties in the GC~\cite{calore2014} has been placed under some pressure by existing data~\cite{gs2016}, with great promise shown by the Square Kilometre Array~\cite{ska} (SKA) as a future probe of these models.

A renewed reason for interest in the GC excess for DM hunting is that a similar spectrum of gamma-rays has been observed by Fermi-LAT in the M31 galaxy~\cite{fermim312017}, which is expected to be followed up by a detailed analysis on the possibility of DM contributing to the spectrum~\cite{fermihawc2017}. Therefore, as a complement to this, it is of interest to examine how DM annihilation cross-section constraints, derived from existing independent astrophysical data, could impact on a possible DM contribution to the gamma-ray spectrum of M31. It is also important to determine whether or not a DM contribution to M31 gamma-rays would be consistent with the models contributing to the excess of gamma-rays in the GC, as the gamma-ray spectra of these two targets are morphologically similar, both being confined to the inner region of the galaxies, and suggest a similar non-inter-stellar origin~\cite{fermim312017}.

In this work we will determine the largest fraction of the total integrated gamma-ray fluxes from M31 and the GC that can be accounted for by DM, given existing data on the Coma galaxy cluster (diffuse radio~\cite{coma-radio2003} and gamma-rays~\cite{fermicoma2015}) and the Reticulum II dwarf galaxy gamma-ray limits~\cite{Fermidwarves2015}. These sources were chosen due to the availability of spectral data as well as their ability to provide good constraints on the DM parameter space~\cite{Colafrancesco2006,Fermidwarves2015,gs2016}. In order to make a direct comparison between M31 and the GC we will confine our attention to the region of the gamma-ray spectrum extending from $1$ GeV to $12$ GeV using the data sets from \cite{fermigc2015,fermim312017}. This is in order to encompass the region of the high-energy gamma-ray excess observed by Fermi-LAT~\cite{fermigc2015} in the GC and accommodate the more limited M31 data set. Above $12$ GeV, only upper-limits exist on the spectrum of M31, with these being greatly in excess of any power-law trend fitting the preceding data points~\cite{fermim312017}. Since this is very different to the GC measured spectrum we will omit higher energies in order to make direct comparison between these two targets. We note that results obtained by looking bin by bin for the maximum ratio between DM and observed fluxes do not differ substantially from those presented here.

We find that a vital factor in determining how significant the DM contribution is to these galactic gamma-ray spectra is the assumed density profile of the DM halo. We compare both a canonical Navarro-Frenk-White (NFW) profile~\cite{nfw1996} and a contracted NFW profile with a similar exponent to that used to fit DM models to the GC excess~\cite{daylan2016}. This reveals that, in order for the DM contribution to be significant ($\simgt \mathcal{O}(10\%)$), a contracted cuspy halo is necessary. This is important, as there is evidence to suggest that both the Milky-Way and M31 data prefer cored halos~\cite{nesti2013,banerjee}, as well as anything smaller than a galaxy cluster~\cite{rodrigues2017}. Thus, if the halo profiles of these targets prove to be cored, or even fail to be steeply cusped, then both gamma-ray and radio data from Coma and Reticulum II will constrain the DM contribution to the M31 and GC gamma-ray spectra to, or below, the level of $1\%$ for the studied energy range.

This paper is structured as follows: Section~\ref{sec:ann} reviews the DM annihilation formalism, with the resulting emissions being discussed in Section~\ref{sec:emm}. The models that will be used for the various DM halos are detailed in Section~\ref{sec:halos} and the derivation of the cross-section constraints from Coma and Reticulum II is explained in Section~\ref{sec:cx}. Finally, the results are presented in Section~\ref{sec:res} and discussed in \ref{sec:conc}.

\section{Dark Matter Annihilation}
\label{sec:ann}

The source function for particle $i$ (electrons/positrons or photons) with energy $E$ from a DM annihilation is taken to be
\begin{equation}
Q_i (r,E) = \langle \sigma V\rangle \sum\limits_{f}^{} \td{N^f_i}{E}{} B_f \left(\frac{\rho_{\chi}(r)}{m_{\chi}}\right)^2 \; ,
\end{equation}
where $r$ is distance from the halo centre, $\langle \sigma V \rangle$ is the non-relativistic velocity-averaged annihilation cross-section, $f$ labels the annihilation channel intermediate state with a branching fraction $B_f$ and differential $i$-particle yield $\td{N^f_i}{E}{}$, $\rho_{\chi}(r)$ is the DM radial density profile, and $m_{\chi}$ is the WIMP mass. The $f$ channels used will be quarks $q\bar{q}$, electron-positron $e^+e^-$, muons $\mu^+ \mu^-$, $\tau$-leptons $\tau^+\tau^-$, Higgs bosons $hh$, $W$ bosons $W^+W^-$, $Z$ bosons $ZZ$, and photons $\gamma\gamma$.

We will treat each annihilation channel separately, setting $B_f = 1$ for the channel of interest in each case. The yield functions $\td{N^f_i}{E}{}$ are taken from \cite{ppdmcb1,ppdmcb2} for all channels (with electro-weak corrections), however, when $m_{\chi} < (m_Z , \, m_W)$ the model independent formulation within the micrOMEGAs package~\cite{micromegas1,micromegas2} is used instead for the $ZZ$ and $W^+W^-$ channels.

\section{Dark Matter Induced Emissions}
\label{sec:emm}

The average power of the synchrotron radiation at observed frequency $\nu$ emitted by an electron with energy $E$ in a magnetic field with amplitude $B$ is given by~\cite{longair1994}
\begin{equation}
P_{synch} (\nu,E,r,z) = \int_0^\pi d\theta \, \frac{\sin{\theta}^2}{2}2\pi \sqrt{3} r_e m_e c \nu_g F_{synch}\left(\frac{\kappa}{\sin{\theta}}\right) \; ,
\label{eq:power}
\end{equation}
where $m_e$ is the electron mass, $\nu_g = \frac{e B}{2\pi m_e c}$ is the non-relativistic gyro-frequency, $r_e = \frac{e^2}{m_e c^2}$ is the classical electron radius, and the quantities $\kappa$ and $F_{synch}$ are defined as
\begin{equation}
\kappa = \frac{2\nu (1+z)}{3\nu_g \gamma^2}\left[1 +\left(\frac{\gamma \nu_p}{\nu (1+z)}\right)^2\right]^{\frac{3}{2}} \; ,
\end{equation}
with $\nu_p \propto \sqrt{n_e}$, and
\begin{equation}
F_{synch}(x) = x \int_x^{\infty} dy \, K_{5/3}(y) \simeq 1.25 x^{\frac{1}{3}} \mbox{e}^{-x} \left(648 + x^2\right)^{\frac{1}{12}} \; .
\end{equation}
The average power produced by inverse-Compton Scattering (ICS) of a low energy photon distribution is given by
\begin{equation}
P_{IC} (\nu,E,z) = c E_{\gamma}(z) \int d\epsilon \; n(\epsilon) \sigma(E,\epsilon,E_{\gamma}(z)) \; ,
\label{eq:ics_power}
\end{equation}
where $E_{\gamma}(z) = h \nu (1+z)$ is the emitted photon energy, $n(\epsilon)$ is the black-body spectrum of the CMB photons, and $E$ is the electron energy. Here we consider only the ICS of CMB photons, because this is the largest radiation background available in the universe.
Additionally,
\begin{equation}
\sigma(E,\epsilon,E_{\gamma}) = \frac{3\sigma_T}{4\epsilon\gamma^2}G(q,\Gamma_e) \; ,
\end{equation}
where $\sigma_T$ is the Thompson cross-section, $\gamma$ is the electron Lorentz factor, and
\begin{equation}
G(q,\Gamma_e) = 2 q \ln{q} + (1+2 q)(1-q) + \frac{(\Gamma_e q)^2(1-q)}{2(1+\Gamma_e q)} \; ,
\end{equation}
with
\begin{equation}
\begin{aligned}
q & = \frac{E_{\gamma}}{\Gamma_e(\gamma m_e c^2 + E_{\gamma})} \; , \\
\Gamma_e & = \frac{4\epsilon\gamma}{m_e c^2}
\end{aligned}
\end{equation}

Bremsstrahlung emission from secondary electrons produced by DM in the background atmosphere of an astrophysical structure (usually the inter-stellar/galactic or intra-cluster media) has an average power $P_B$ given by
\begin{equation}
P_B (E_{\gamma},E,r) = c E_{\gamma}(z)\sum\limits_{j} n_j(r) \sigma_B (E_{\gamma},E) \; ,
\end{equation}
where $n_j(r)$ is the density of the background atmosphere species labelled $j$, and
\begin{equation}
\sigma_B (E_{\gamma},E) = \frac{3\alpha \sigma_T}{8\pi E_{\gamma}}\left[ \left(1+\left(1-\frac{E_{\gamma}}{E}\right)^2\right)\phi_1 - \frac{2}{3}\left(1-\frac{E_{\gamma}}{E}\right)\phi_2 \right] \; ,
\end{equation}
with $\phi_1$ and $\phi_2$ being energy dependent factors determined by the species $j$(see \cite{longair1994}).

For the DM-induced $\gamma$-ray production, the flux calculation is somewhat simplified
\begin{equation}
S_{\gamma} (\nu,z) = \int_0^r d^3r^{\prime} \, \frac{Q_{\gamma}(\nu,z,r)}{4\pi D_L^2} \; ,
\end{equation}
with $Q_{\gamma}(\nu,z,r)$ being the source function within the given DM halo.

The local emissivity for the $i-th$ emission mechanism (synchrotron, ICS, bremsstrahlung) can then be found as a function of the electron and positron equilibrium distributions as well as the associated power
\begin{equation}
j_{i} (\nu,r,z) = \int_{m_e}^{M_\chi} dE \, \left(\td{n_{e^-}}{E}{} + \td{n_{e^+}}{E}{}\right) P_{i} (\nu,E,r,z) \; ,
\label{eq:emm}
\end{equation}
where $\td{n_{e^-}}{E}{}$ is the equilibrium electron distribution from DM annihilation (see below).
The flux density spectrum within a radius $r$ is then written as
\begin{equation}
S_{i} (\nu,z) = \int_0^r d^3r^{\prime} \, \frac{j_{i}(\nu,r^{\prime},z)}{4 \pi D_L^2} \; ,
\label{eq:flux}
\end{equation}
where $D_L$ is the luminosity distance to the halo.  

In the case of Reticulum II, the GC, and M31 we will instead calculate the resulting gamma-ray flux based on the astrophysical J-factor, described below in Section~\ref{sec:halos},
\begin{equation}
S_{\gamma} (\nu,z) = \langle \sigma V\rangle \sum\limits_{f}^{} \td{N^f_i}{E}{} B_f J(\Delta \Omega,l) \; ,
\end{equation}
this form will be used for all spectra relevant to Reticulum II, GC, and M31 rather than Eq.~(\ref{eq:flux}). Note that this means we are not considering gamma-ray emission from secondary electron dependent processes (ICS and Bremsstrahlung). Since the M31 and GC gamma-ray data is binned~\cite{fermim312017,fermigc2015}, the DM gamma-ray flux spectrum $S$ will also be binned for accurate comparison. 

We stress that the fluxes for Coma are treated differently in order for the radio and gamma-ray spectra to be calculated in a consistent manner.

In electron-dependent emissions there are two important processes that effect the energy and spatial distribution of DM-produced electrons, namely energy-loss and diffusion. Diffusion is typically only significant within small structures~\cite{Colafrancesco2007,gsp2015}, thus it will not be relevant to the Coma cluster (which is the only case where we perform these calculations - as we do not consider electron-based emission processes in Reticulum II, M31, or the GC).
The equilibrium electron distribution is found as a stationary solution to the equation
\begin{equation}
\begin{aligned}
\pd{}{t}{}\td{n_e}{E}{} = & \; \gv{\nabla} \left( D(E,\v{r})\gv{\nabla}\td{n_e}{E}{}\right) + \pd{}{E}{}\left( b(E,\v{r}) \td{n_e}{E}{}\right) + Q_e(E,\v{r}) \; ,
\end{aligned}
\end{equation}
where $D(E,\v{r})$ is the diffusion coefficient, $b(E,\v{r})$ is the energy loss function, and $Q_e(E,\v{r})$ is the electron source function from DM annihilation. In this case, we will work under the simplifying assumption that $D$ and $b$ lack a spatial dependence and thus we will include only average values for magnetic field and thermal electron densities. The solution for the Coma cluster, when diffusion is negligible, has the form
\begin{equation}
\td{n_e}{E}{} = \frac{1}{b(E)} \int_E^{m_\chi} \, dE^{\prime} \, Q_e (r, E^{\prime}) \; .
\end{equation}
For details of the solution see \cite{Colafrancesco2007}. 
%

The energy loss function is defined by
\begin{equation}
\begin{aligned}
b(E) = & b_{IC} E^2 (1+z)^4 + b_{sync} E^2 \overline{B}^2 \\&\; + b_{coul} \overline{n} (1+z)^3 \left(1 + \frac{1}{75}\log\left(\frac{\gamma}{\overline{n} (1+z)^3}\right)\right) \\&\; + b_{brem} \overline{n} (1+z)^3 \left( \log\left(\frac{\gamma}{\overline{n} (1+z)^3 }\right) + 0.36 \right) \;,
\end{aligned}
\label{eq:loss}
\end{equation}
where $\overline{n}$ is the average thermal electron density in the halo and is given in cm$^{-3}$, $\overline{B}$ is the average magnetic field in $\mu$G, while $b_{IC}$, $b_{sync}$, $b_{coul}$, and $b_{brem}$ are the inverse Compton, synchrotron, Coulomb and bremsstrahlung energy loss factors, taken to be $0.25$, $0.0254$, $6.13$, and $1.51$ respectively in units of $10^{-16}$ GeV s$^{-1}$. Here $E$ is the energy in GeV.

\section{Dark Matter Halos}
\label{sec:halos}

For the Coma cluster DM halo we consider the model described in~\cite{Colafrancesco2006}.
The virial mass of this cluster is taken to be $M_{vir} = 1.33 \times 10^{15}$ M$_{\odot}$, with virial concentration  $c_{vir} = 10$, at the redshift $z = 0.0231$.
The density profile used in the NFW~\cite{nfw1996} form
\begin{equation}
\rho(r)=\frac{\rho_s}{\frac{r}{r_s}\left(1+\frac{r}{r_s}\right)^{2}} \; ,
\label{eq:nfw}
\end{equation}
with $r_s = \frac{R_{vir}}{c_{vir}}$ being the scale radius of the profile, and $\rho_s$ is the halo characteristic density appropriately normalised to $M_{vir}$.
This smooth profile is supplemented by the effects of sub-halos following the method described in \cite{prada2013} (a similar process is described in \cite{ng2014}), this results in a boosting of the DM-induced flux by a factor of $\sim 30$. The Coma cluster is the only environment within which we will make use of a boosting factor.

The thermal electron density of the Intra-Cluster Medium (ICM) in Coma $n_e(r)$ is given by~\cite{briel1992}
\begin{equation}
n_e(r) = n_0 \left(1 + \left[\frac{r}{r_s}\right]^2\right)^{-q_e} \; ,
\end{equation}
with $r_s$ being a characteristic radius (taken equal to the halo scale radius), $n_0 = 3.44 \times 10^{-3}$ cm$^{-3}$ and $q_e = 1.125$~\cite{briel1992}.
The magnetic field in Coma is assumed to follow the one derived by \cite{bonafede2010} having a radial profile given by
\begin{equation}
B(r) = B_0 \left(\frac{n_e(r)}{n_0}\right)^{q_b} \; ,
\end{equation}
where $r$ is the distance from the cluster centre, $B_0 = 4.7$ $\mu$G, and $q_b = 0.5$.
Additionally, this magnetic field has a Kolmogorov turbulence power spectrum with a minimal coherence length of $\approx 2$ kpc.\\

For comparison of Reticulum II gamma-ray data cross-section limits to the spectra of M31 and the Galactic Centre we will make use of the astrophysical J-factor to describe the halos (in the case of Coma we calculate emissions directly from the halo profile for consistency with radio emissions)
\begin{equation}
J (\Delta \Omega, l) = \int_{\Delta \Omega}\int_{l} \rho^2 (\v{r}) dl^{\prime}d\Omega^{\prime} \; , \label{eq:jfactor}
\end{equation}
with the integral being extended over the line of sight $l$, and $\Delta \Omega$ is the observed solid angle. All $J$-factors for the various halos studied are summarised in the Table~\ref{tab:jfactors}. 

In the calculation of J-factors (for the GC and M31) we will consider both the NFW halo profile from Eq.~(\ref{eq:nfw}) as well as the contracted NFW profile
\begin{equation}
\rho(r)=\frac{\rho_s}{\left(\frac{r}{r_s}\right)^\gamma\left(1+\frac{r}{r_s}\right)^{3-\gamma}} \; ,
\label{eq:nfwcon}
\end{equation}
with $\gamma = 1.3$ for comparison to GC best-fit DM models~\cite{daylan2016}.
In the Galactic Centre we follow \cite{fermigc2015} in using the J-factor for the observed region to be $2\times 10^{22}$ GeV$^2$ cm$^{-5}$ for a NFW profile, and $J \sim 4 \times 10^{23}$ GeV$^2$ cm$^{-5}$ for a contracted NFW profile following the method used in \cite{daylan2016} (and matching their best-fit DM models). While, in M31, we follow the results quoted in \cite{fermim312017} of $J \sim 8 \times 10^{18}$ GeV$^2$ cm$^{-5}$~\cite{fermihawc2017} for the region observed to emit in gamma-rays in an NFW halo ($\sim 1 \times 10^{20}$ will be used for the case of a contracted NFW halo following the same method as above). It is important to note that we do not employ any boosting factor for substructure on top of these J-factors. 

\begin{table}
\begin{tabular}{|l|l|l|}
\hline
Halo & $J_{\gamma=1}$ (GeV$^2$ cm$^{-5}$) & $J_{\gamma=1.3}$ (GeV$^2$ cm$^{-5}$) \\
\hline
GC & $2\times 10^{22}$ & $4 \times 10^{23}$ \\
M31 & $8\times 10^{18}$ & $1 \times 10^{20}$ \\
\hline
\end{tabular}
\begin{tabular}{|l|l|}
\hline
Halo & $J$ (GeV$^2$ cm$^{-5}$) \\
\hline
Reticulum II & $2.0 \times 10^{19}$ \\
Coma & $1.0 \times 10^{18}$ \\
\hline
\end{tabular}
\caption{Summary of $J$-factors for DM halos of interest. For M31 and the GC: $J_{\gamma=1}$ is the NFW $J$-factor, while $J_{\gamma=1.3}$ is that of the contracted NFW case.}
\label{tab:jfactors}
\end{table}

For the Reticulum II dwarf we take the J-factor to be $2.0 \times 10^{19}$ GeV$^2$ cm$^{-5}$ with a systematic relative error factor of $\sim 1.5$~\cite{Fermidwarves2015}. We note that the estimates for $J$ in \cite{bonnivard2015} are larger over the same angular region but have substantial uncertainties (relative error factor $\sim 10$).

\section{Cross-Section Constraints}
\label{sec:cx}

Using the halo models from Section~\ref{sec:halos} above, we will employ Coma diffuse radio data~\cite{coma-radio2003} as well as gamma-ray data for Coma~\cite{fermicoma2015} and Reticulum II~\cite{Fermidwarves2015} in order to derive $3\sigma$ confidence level upper-limits on the WIMP annihilation cross-section for various annihilation channels and WIMP masses. The cross-section limits found in this manner are equivalent to those from \cite{gs2016}. These cross-sections will then be employed to calculate an expected flux from DM annihilation in the GC and M31 halos, using the limits from Reticulum II and Coma separately, and comparing the resulting integrated fluxes with those from the M31 and GC data~\cite{fermim312017,fermigc2015} for the given range of energies ($1$ to $12$ GeV). The fractional contribution of DM to the integrated flux will then be
\begin{equation}
f_{\chi,max} = \frac{\mathcal{F}_{DM,\mathcal{H}}\langle \sigma V \rangle_{upper}}{\mathcal{F}_{\mathcal{H}}} \; ,
\end{equation} 
where $\mathcal{F}_{DM,\mathcal{H}}$ is the integrated DM spectrum for the halo (GC or M31) $\mathcal{H}$, $\mathcal{F}_{\mathcal{H}}$ is the integrated gamma-ray spectrum for the target halo from the gamma-ray data, and $\langle \sigma V \rangle_{upper}$ are upper limits~\cite{gs2016} taken from either Coma data (radio or gamma-ray) or Reticulum II (gamma-ray only). 

A similar analysis was conducted on the fluxes bin-by-bin, in each bin the DM and known spectra are compared searching for the largest fraction that can be accounted for by the DM emissions. We do not display these results as they do not differ by more than a factor of $2$ from those presented here.

%

\section{Results}
\label{sec:res}

The results presented will consist of the maximum fraction of the total integrated gamma-ray flux that WIMPs of various masses can account for, $f_{\chi,max} (m_{\chi},\langle \sigma V\rangle)$, given the constraints on $\langle \sigma V \rangle$ that can be derived from the Coma cluster and the Reticulum II dwarf data at a confidence level of $3\sigma$. In the interests of direct comparison, we limit the integration region to between $1$ and $12$ GeV to cover the high-energy gamma-ray excess seen in the GC as well as account for the limitations of the M31 data set, where only upper-limits exist above $12$ GeV. For an easier readability, the results for bosonic and fermionic WIMP annihilation channels will be displayed on separate plots in each case.

\begin{figure}[ht!]
\centering
\includegraphics[width=0.45\textwidth]{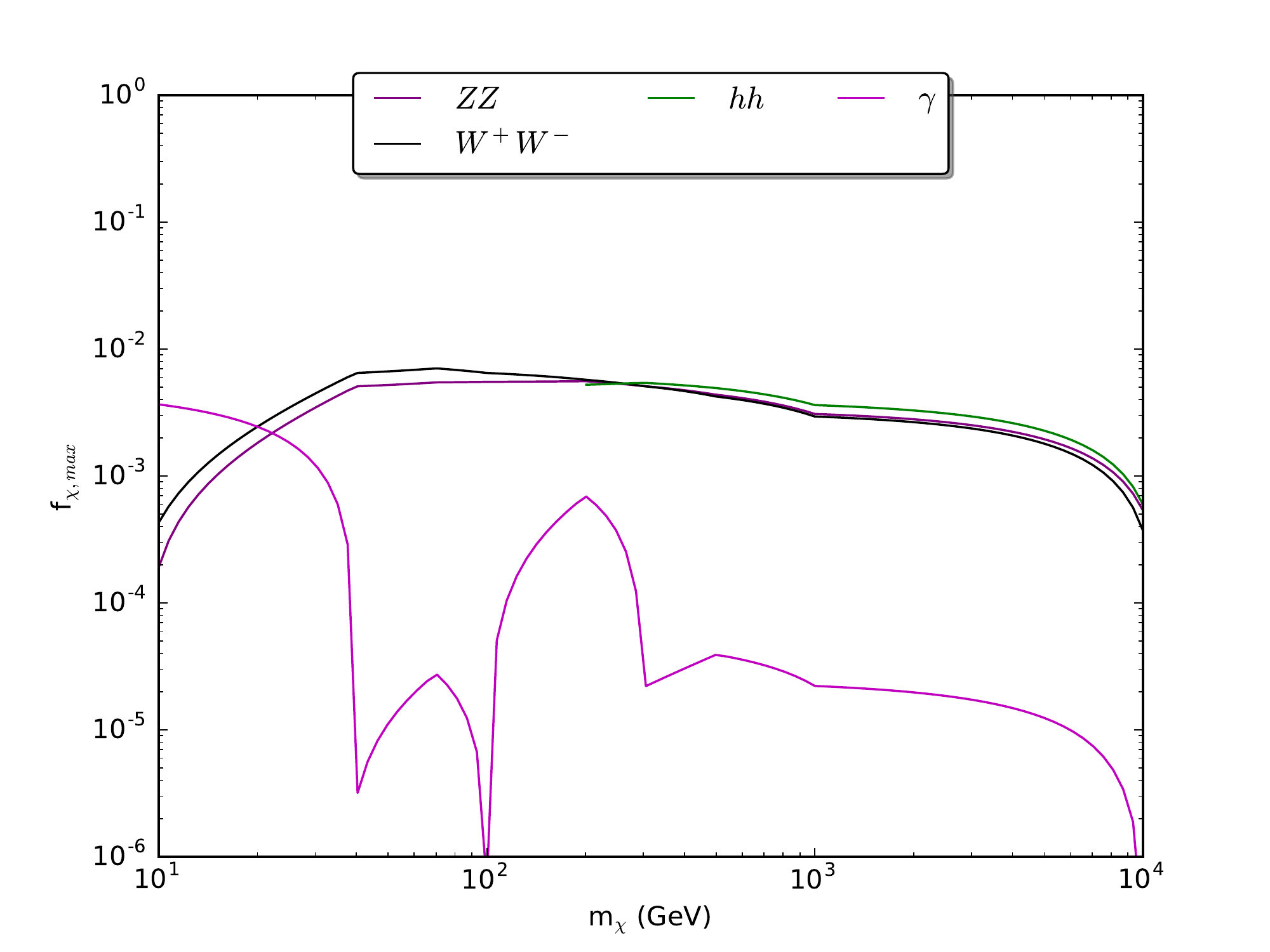}
\includegraphics[width=0.45\textwidth]{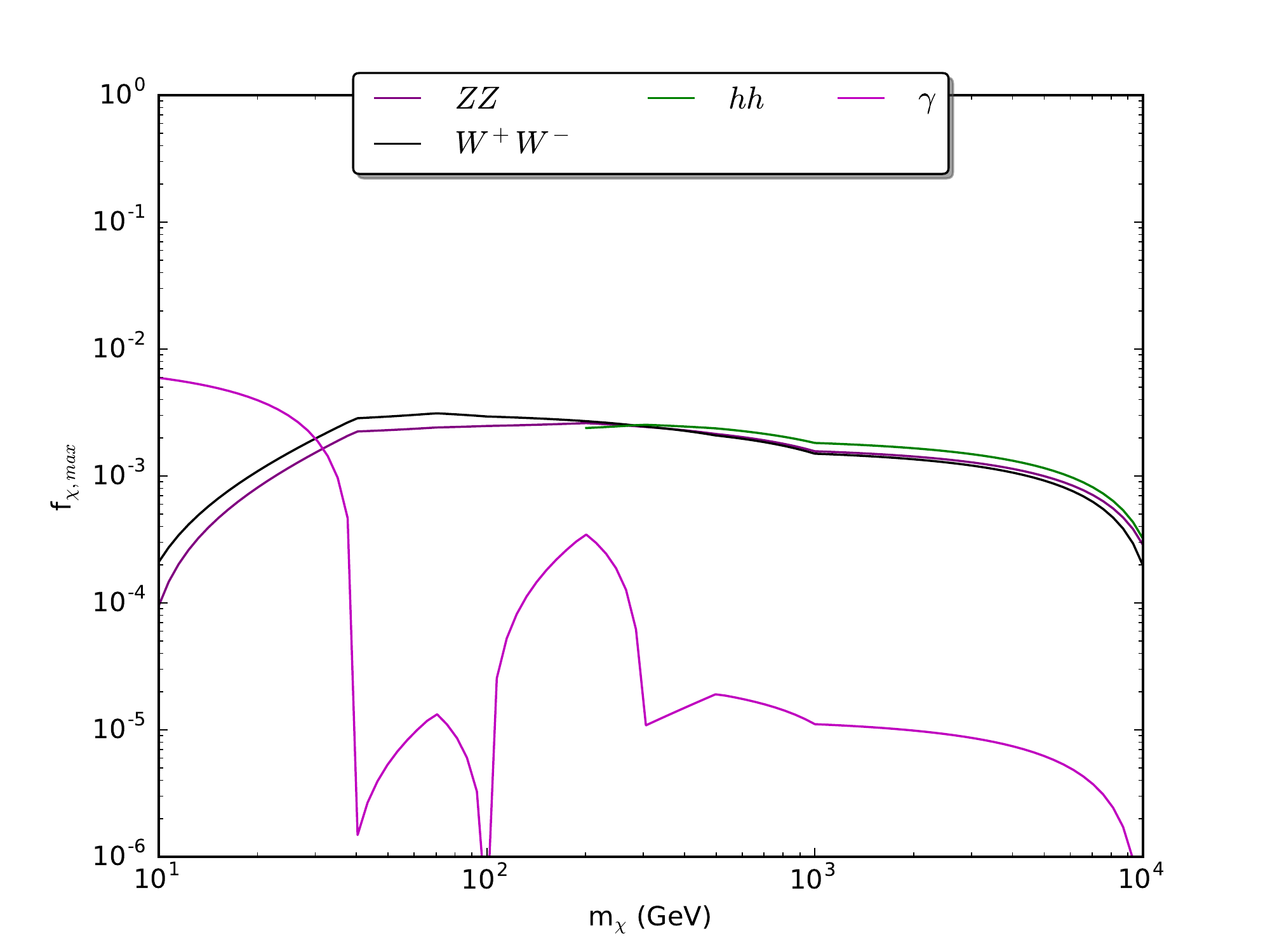}
\caption{Maximum fraction of integrated gamma-ray flux accounted for by DM within M31 and Galactic Centre targets with NFW halos for bosonic channels. Limits are found by imposing cross-section constraints from Coma diffuse radio data~\cite{coma-radio2003}. Left: M31 galaxy. Right: Galactic Centre.}
\label{fig:radiob}
\end{figure}

\begin{figure}[ht!]
\centering
\includegraphics[width=0.45\textwidth]{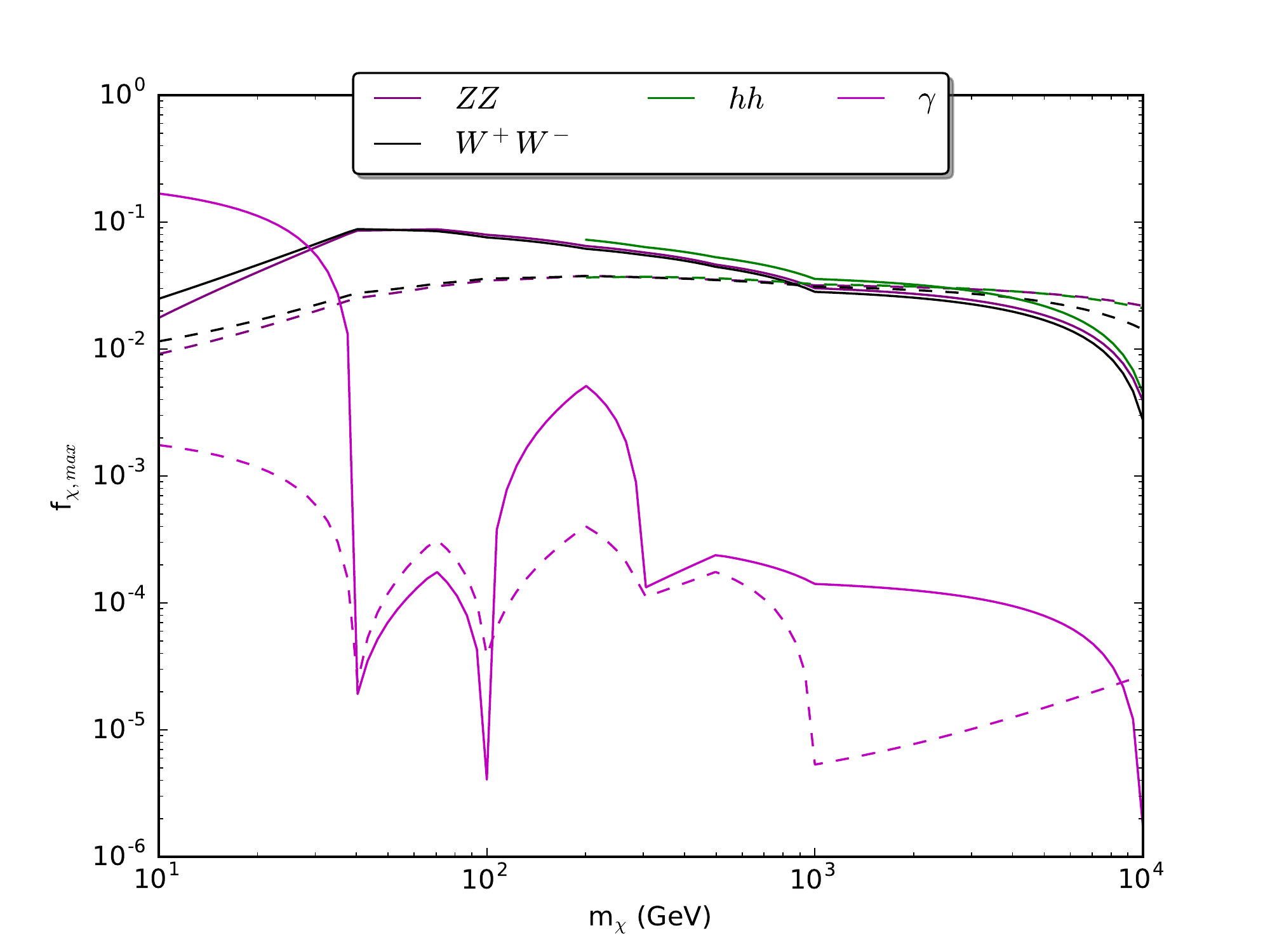}
\includegraphics[width=0.45\textwidth]{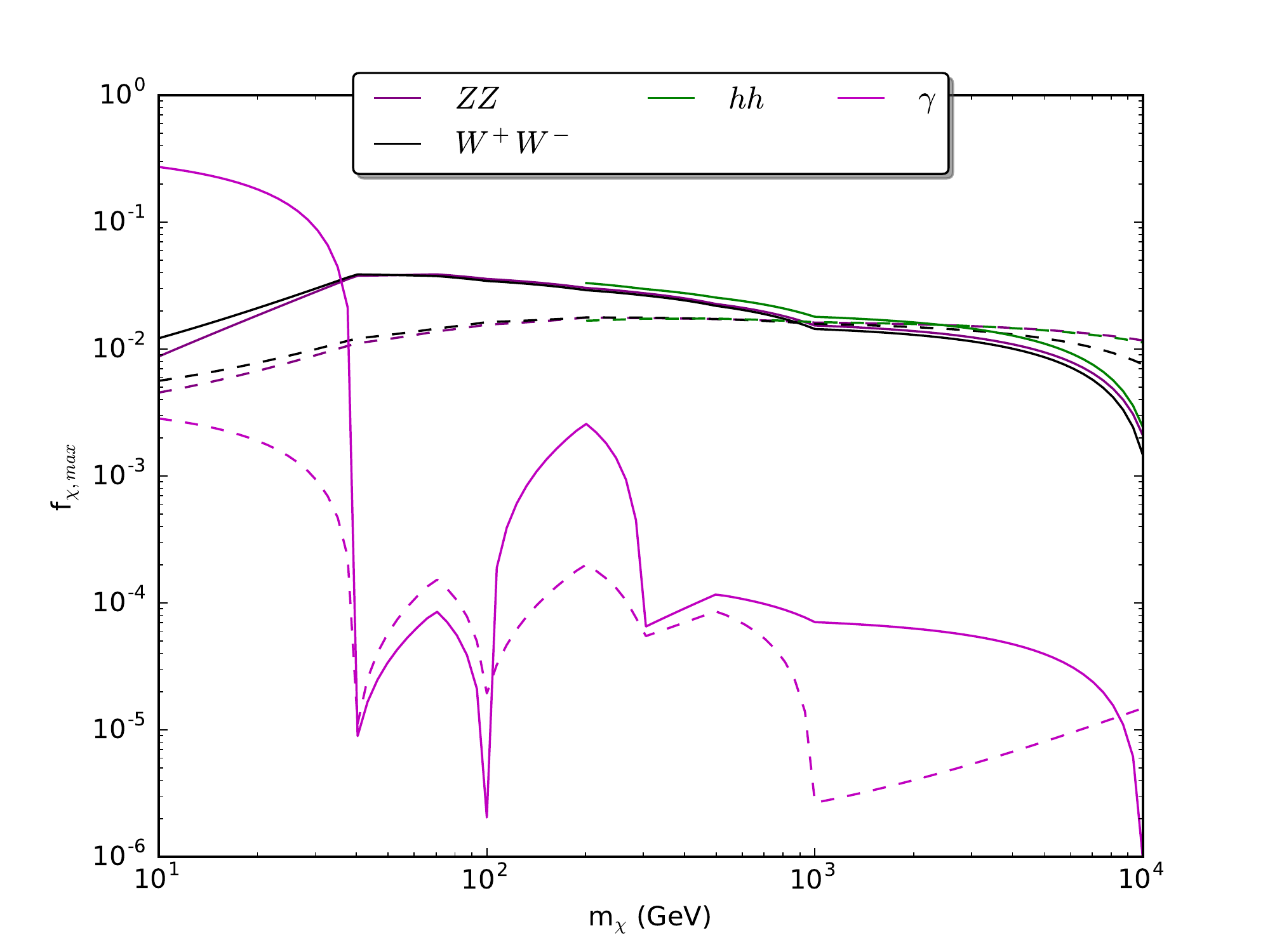}
\caption{Maximum fraction of integrated gamma-ray flux accounted for by DM within M31 and Galactic Centre targets with NFW halos for bosonic channels. Limits are found by imposing cross-section constraints from Fermi-LAT data on Coma~\cite{fermicoma2015} (solid lines) and Reticulum II~\cite{Fermidwarves2015} (dashed lines). Left: M31 galaxy. Right: Galactic Centre.}
\label{fig:gammab}
\end{figure}

\begin{figure}[ht!]
	\centering
	\includegraphics[width=0.45\textwidth]{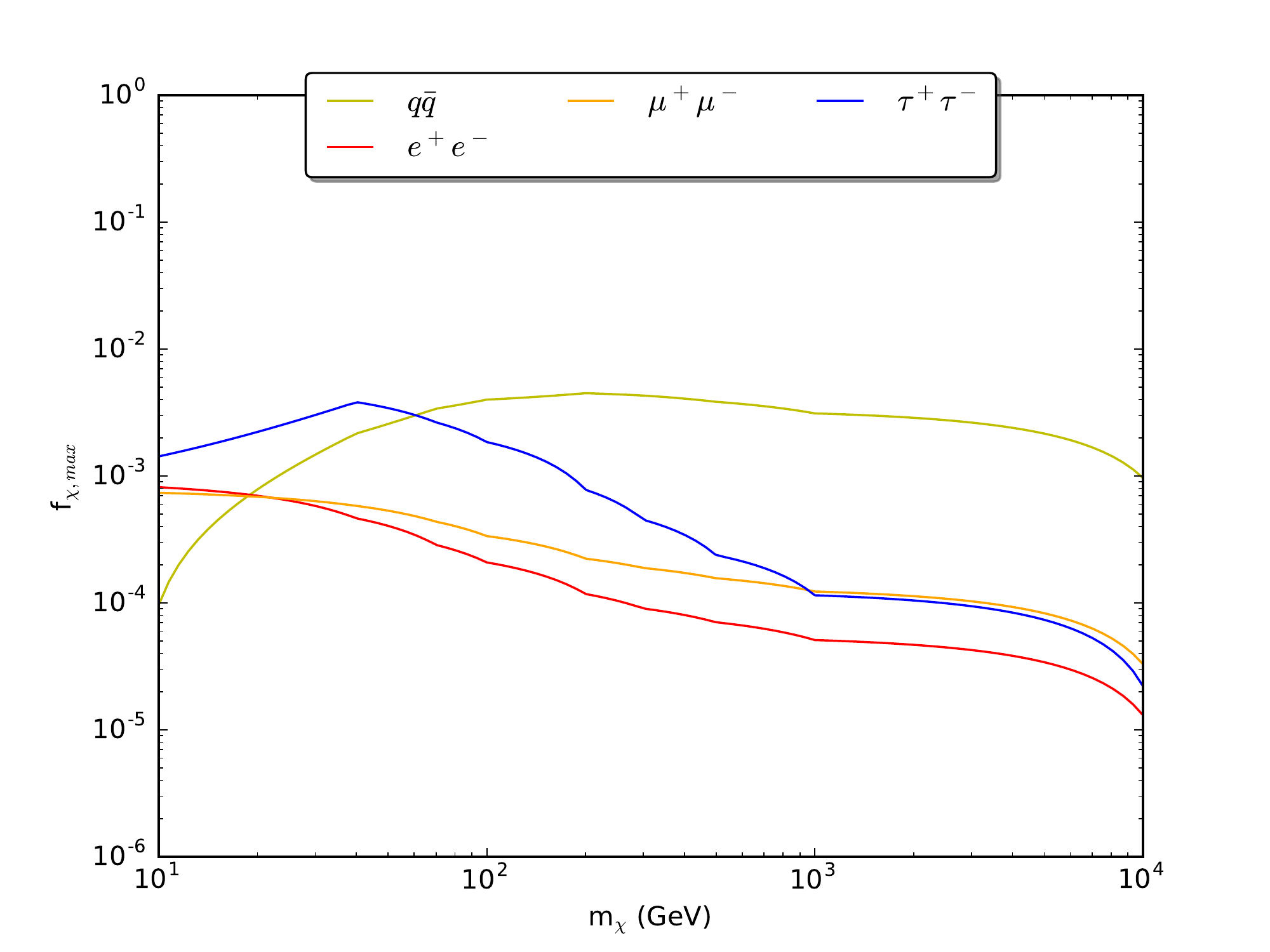}
	\includegraphics[width=0.45\textwidth]{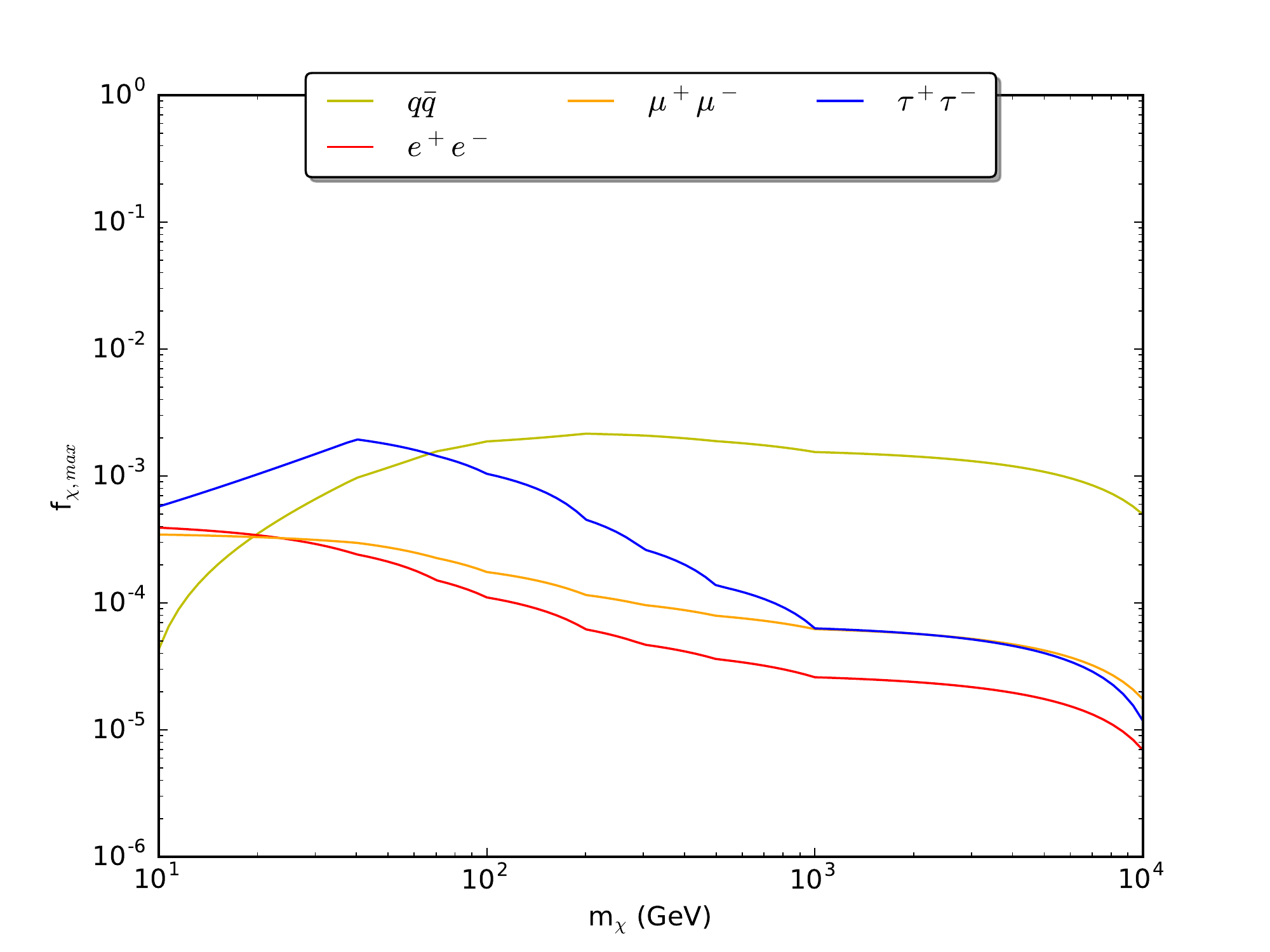}
	\caption{Maximum fraction of integrated gamma-ray flux accounted for by DM within M31 and Galactic Centre targets with NFW halos for fermionic channels. Limits are found by imposing cross-section constraints from Coma diffuse radio data~\cite{coma-radio2003}. Left: M31 galaxy. Right: Galactic Centre.}
	\label{fig:radiof}
\end{figure}

In Figures~\ref{fig:radiob} and \ref{fig:radiof} we show the results produced by imposing WIMP cross-section constraints derived from diffuse Coma radio data~\cite{coma-radio2003} on M31 (left panels) and the Galactic Centre (right panels), when NFW profiles are assumed for both galactic DM halos. What is immediately evident is that the maximal fractions of gamma-rays accounted for by DM in M31 and the Galactic Centre are very similar, being on $\mathcal{O}(1\%)$ at most.  This suggests that DM can account for only a very minor fraction of the observed gamma-rays, far lower indeed than the suggestion briefly sketched out in \cite{fermim312017} of around $20\%$ for M31. The exhaustive nature of the annihilation channels explored ensures that these constraints are comprehensive. It is noteworthy that the two targets show very similar results with the GC having slightly smaller maximum fractions than M31. This is important, as any DM contribution would need to be consistent across the two sources if the gamma-ray spectra are as similar as indicated by \cite{fermim312017}. We also note that these results do not differ by more than a factor of two when compared to the results found by seeking the maximal ratio between the DM and observed spectra bin by bin (this will pertain to all displayed results in this section). It must be noted that these results are sensitive to the DM density profile assumed for Coma, and particularly to the assumptions about substructure within the main halo (this will likely be larger than smooth profile based uncertainties). In addition to this, radio constraints are sensitive to magnetic field uncertainties. In the Coma cluster magnetic field uncertainties amount to $1\sigma$ confidence interval between $3.9$ and $5.4$ $\mu$G about a median of $4.7$ $\mu$G~\cite{bonafede2010}. Thus, we will supplement this data with the gamma-ray limits from the Reticulum II dwarf galaxy where we do not expect significant substructural effects and uncertainties will be sourced from systematics in the $J$-factor estimation~\cite{Fermidwarves2015}.

\begin{figure}[ht!]
	\centering
	\includegraphics[width=0.45\textwidth]{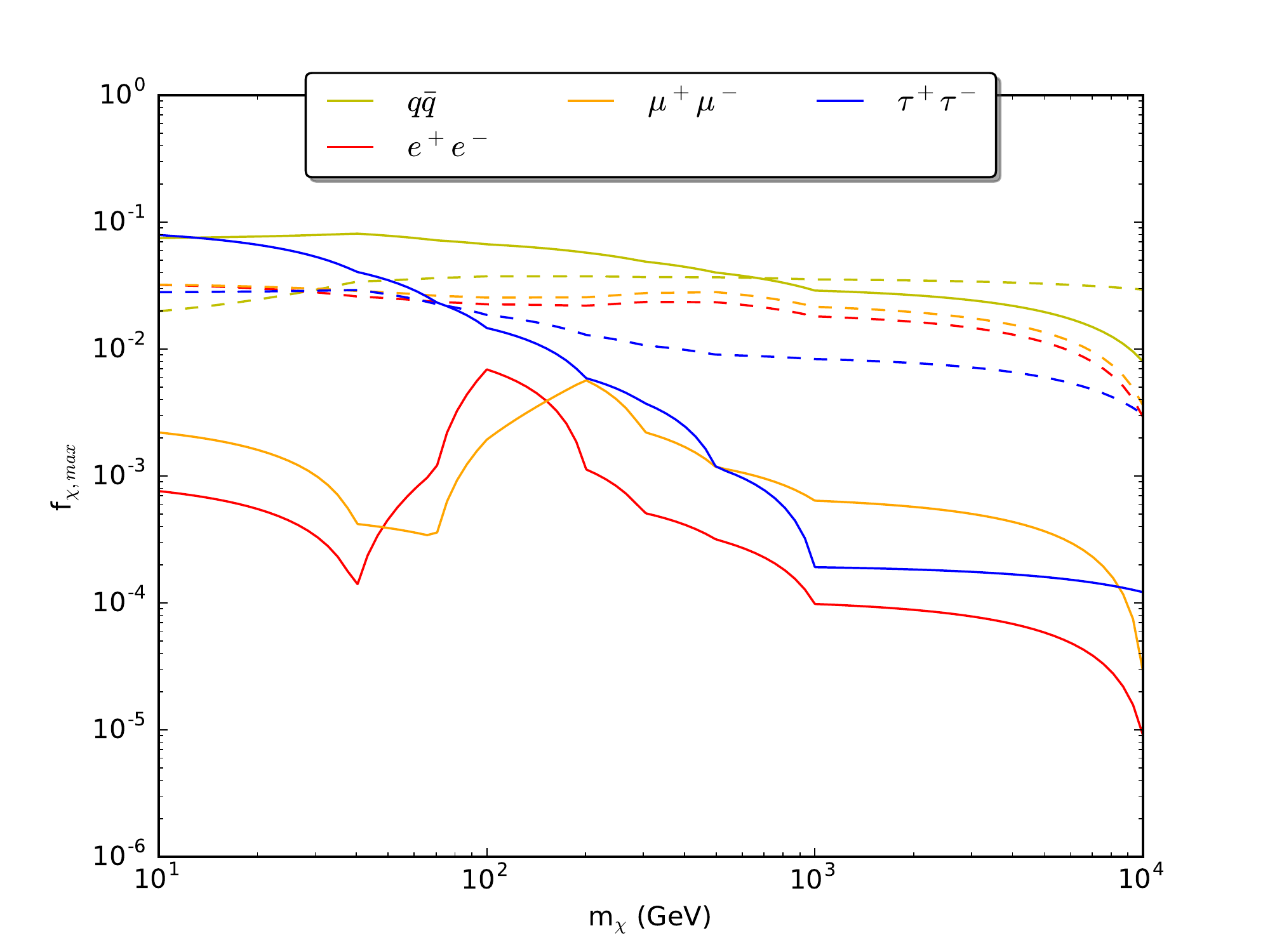}
	\includegraphics[width=0.45\textwidth]{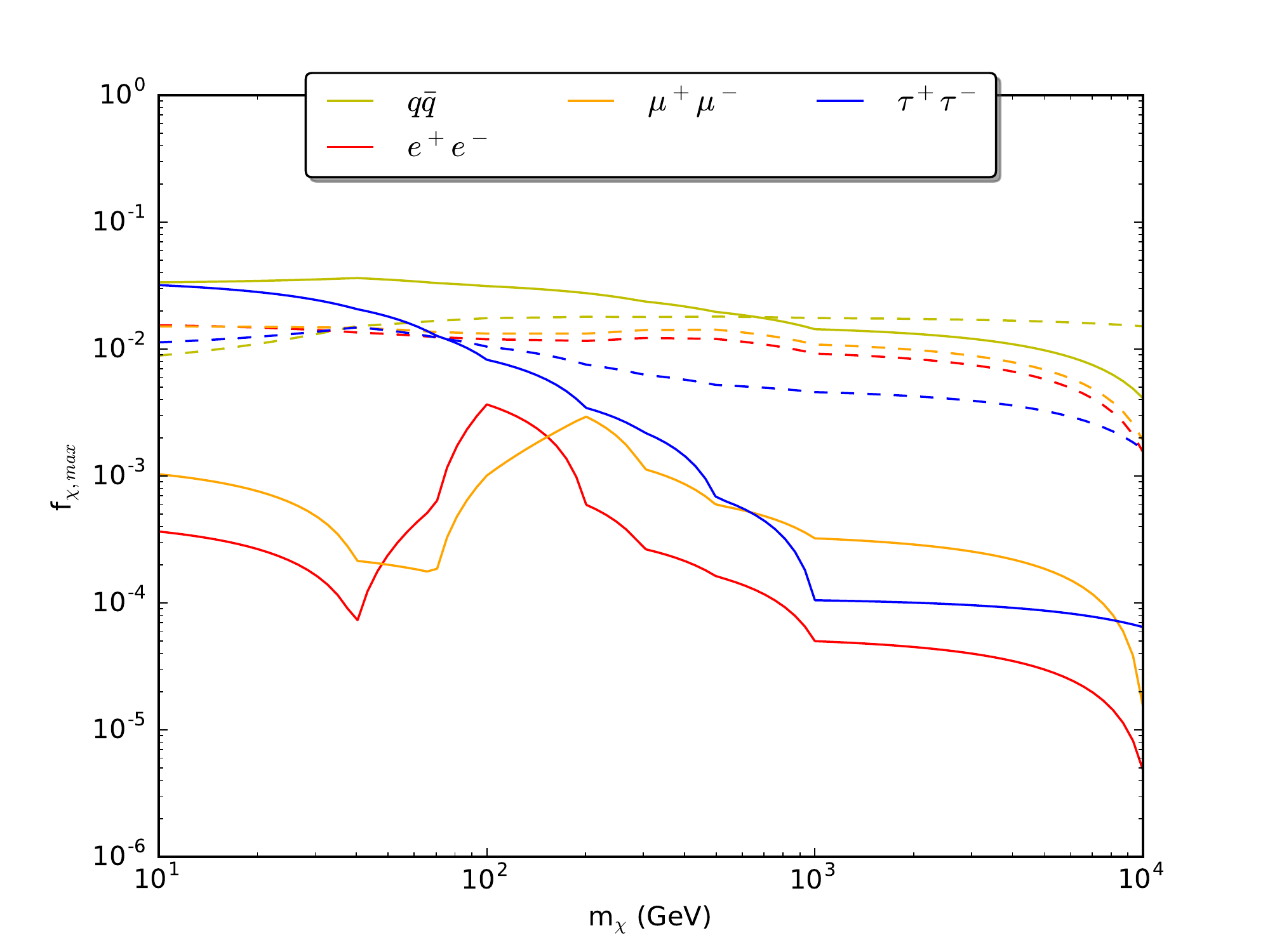}
	\caption{Maximum fraction of integrated gamma-ray flux accounted for by DM within M31 and Galactic Centre targets with NFW halos for fermionic channels. Limits are found by imposing cross-section constraints from Fermi-LAT data on Coma~\cite{fermicoma2015} (solid lines) and Reticulum II~\cite{Fermidwarves2015} (dashed lines). Left: M31 galaxy. Right: Galactic Centre.}
	\label{fig:gammaf}
\end{figure}

In Figures~\ref{fig:gammab} and \ref{fig:gammaf} results analogous to Figs.~\ref{fig:radiob} and \ref{fig:radiof} are shown for the case where our cross-section constraints are drawn from gamma-ray data on Coma and Reticulum II. These limits are considerably more permissive, allowing DM to account for up to $10\%$ of the integrated flux in both targets. Once again, the two targets return very similar maximum DM contributions. It should be noted that the Coma gamma-ray limits are this permissive despite the use of $\sim 30$ substructure boost factor and a full accounting for secondary emissions like bremsstrahlung and ICS. However, in both cases this permissiveness can be reduced in future, as gamma-ray signals in the Coma and Reticulum II targets only have upper-limits on possible emissions.

\begin{figure}[ht!]
\centering
\includegraphics[width=0.45\textwidth]{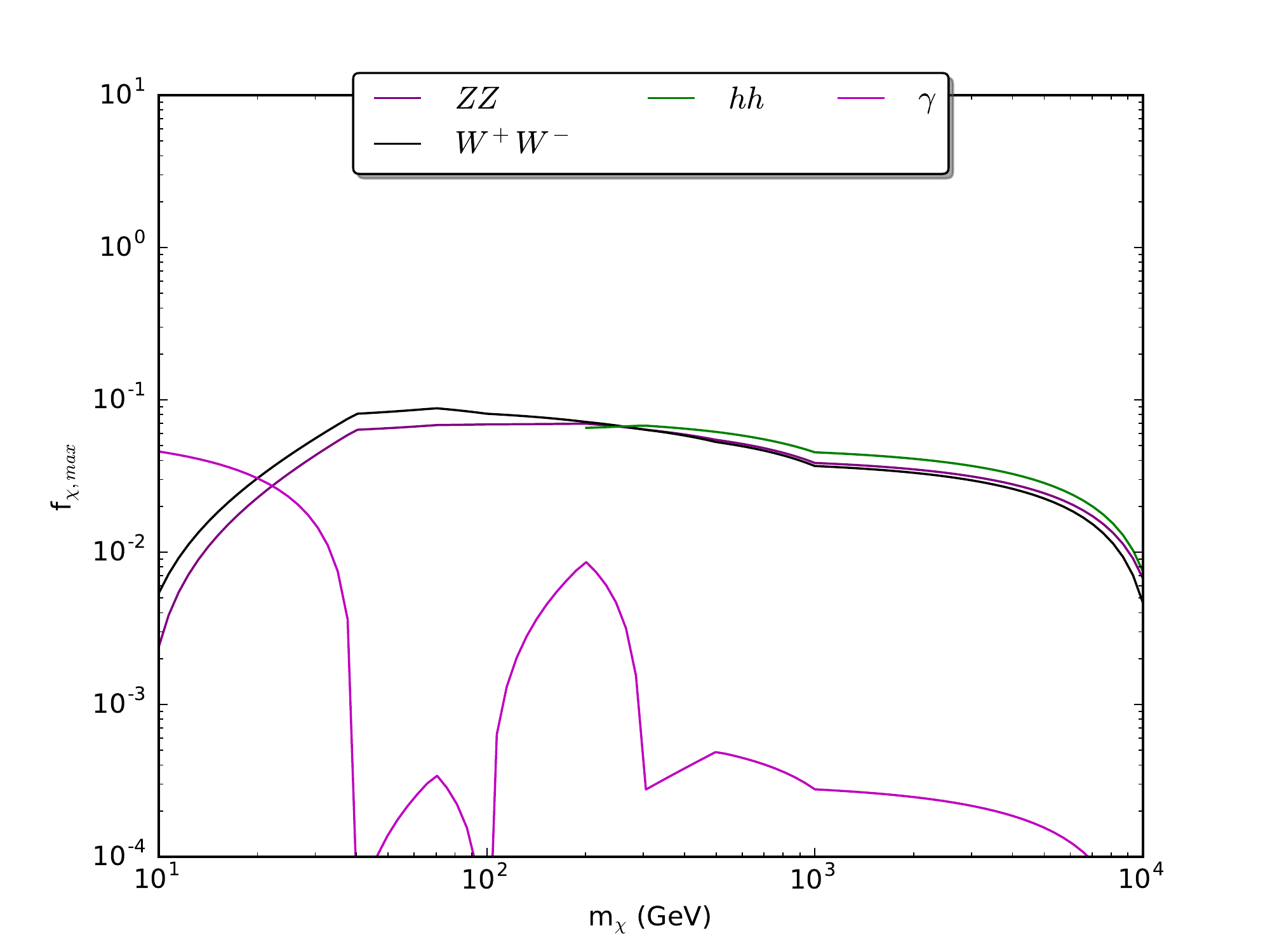}
\includegraphics[width=0.45\textwidth]{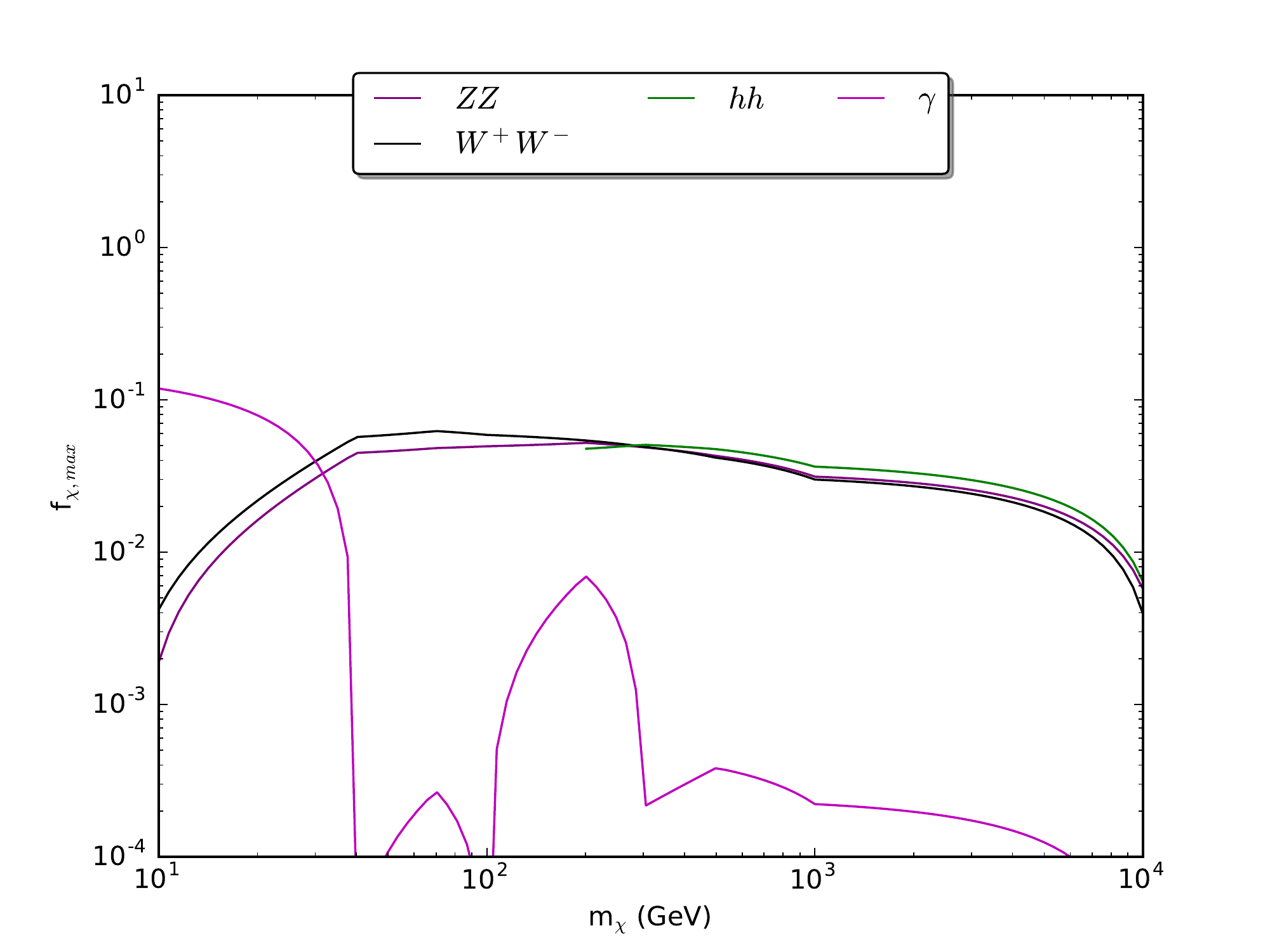}
\caption{Maximum fraction of integrated gamma-ray flux accounted for by DM within M31 and Galactic Centre targets with contracted NFW halos for bosonic channels. Limits are found by imposing cross-section constraints from Coma diffuse radio data~\cite{coma-radio2003}. Left: M31 galaxy. Right: Galactic Centre.}
\label{fig:radio-conb}
\end{figure}

\begin{figure}[ht!]
	\centering
	\includegraphics[width=0.45\textwidth]{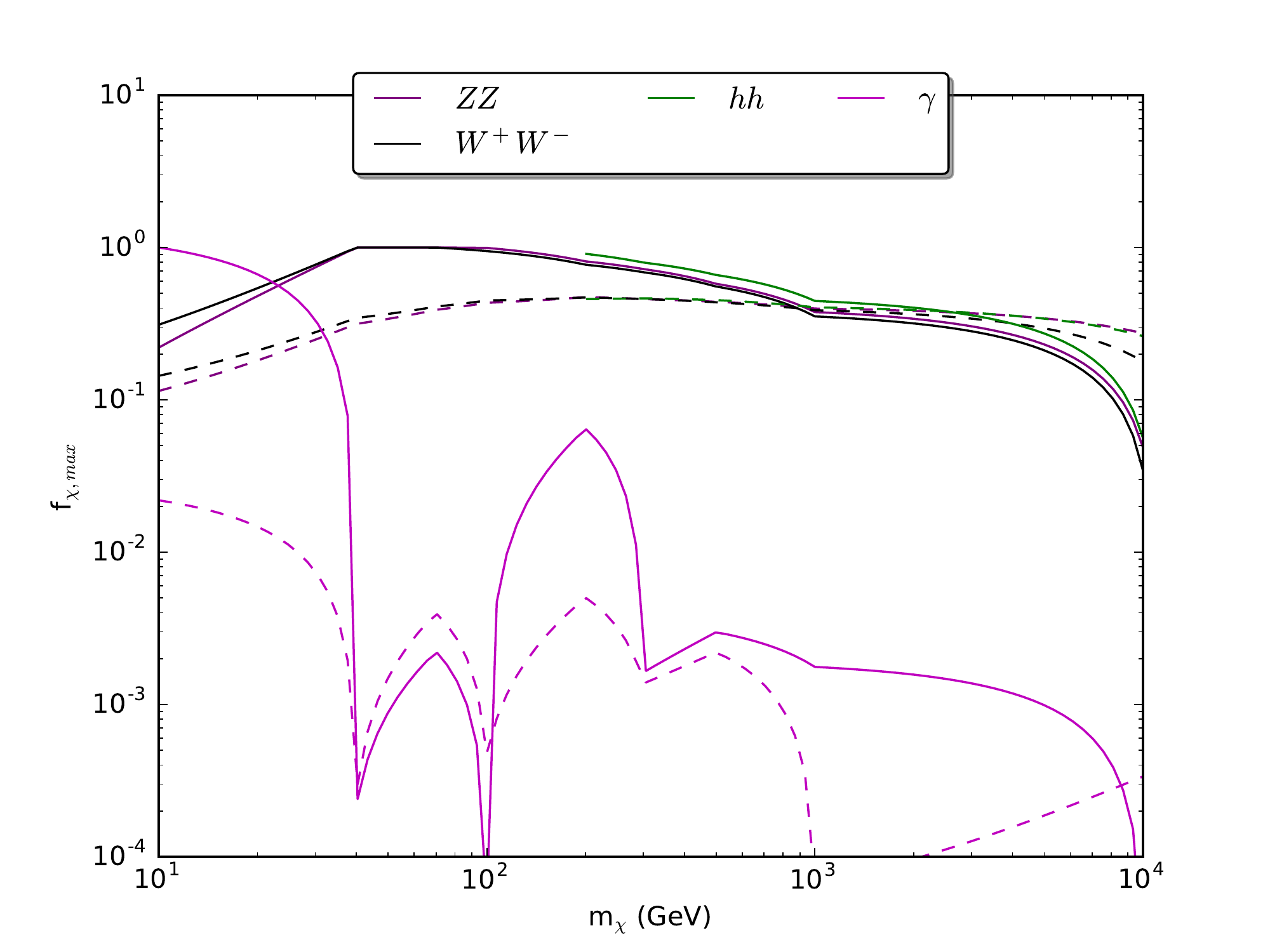}
	\includegraphics[width=0.45\textwidth]{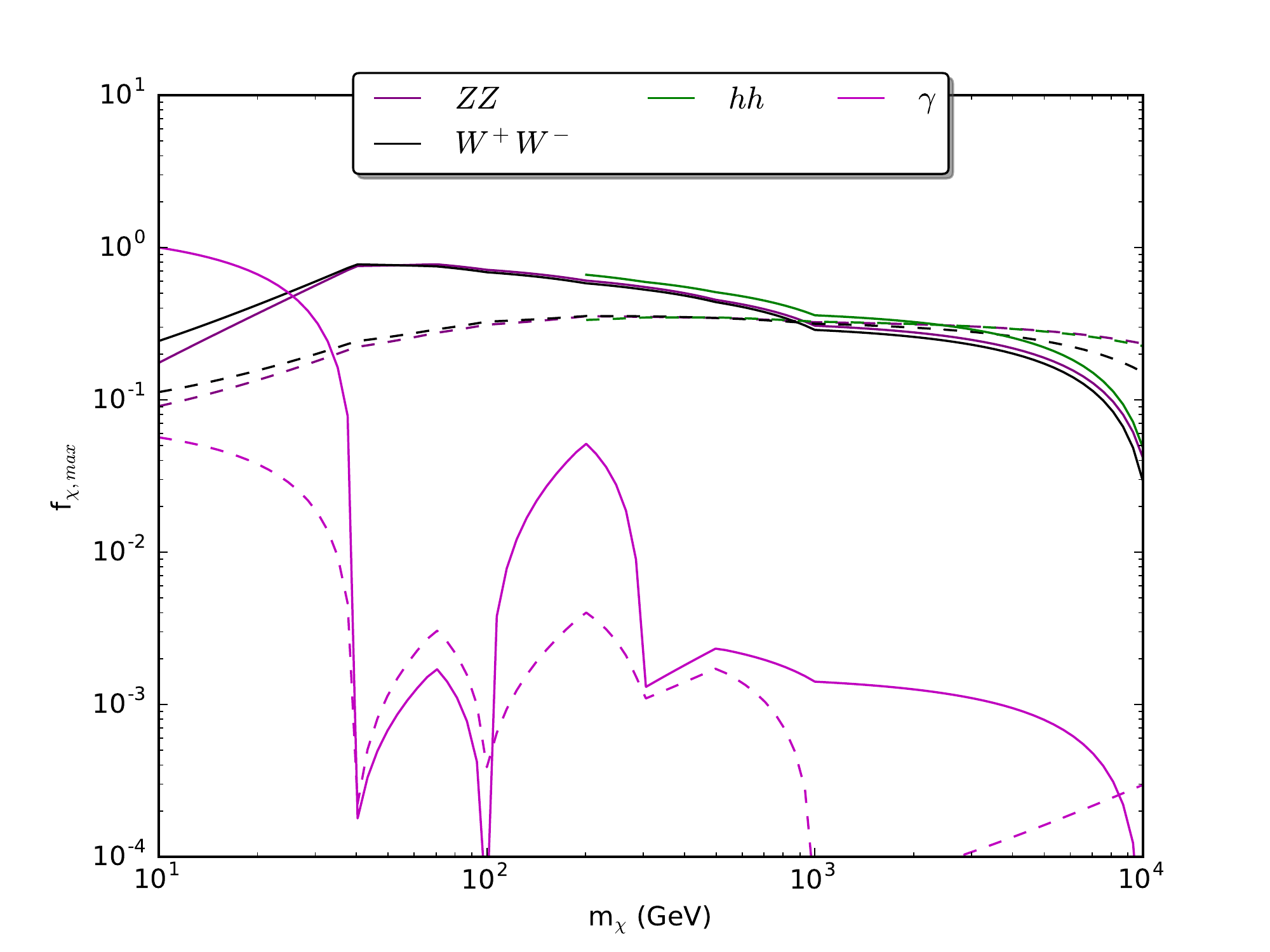}
	\caption{Maximum fraction of integrated gamma-ray flux accounted for by DM within M31 and Galactic Centre targets with contracted NFW halos for bosonic channels. Limits are found by imposing cross-section constraints from Fermi-LAT data on Coma~\cite{fermicoma2015} (solid lines) and Reticulum II~\cite{Fermidwarves2015} (dashed lines). Left: M31 galaxy. Right: Galactic Centre.}
	\label{fig:gamma-conb}
\end{figure}

\begin{figure}[ht!]
	\centering
	\includegraphics[width=0.45\textwidth]{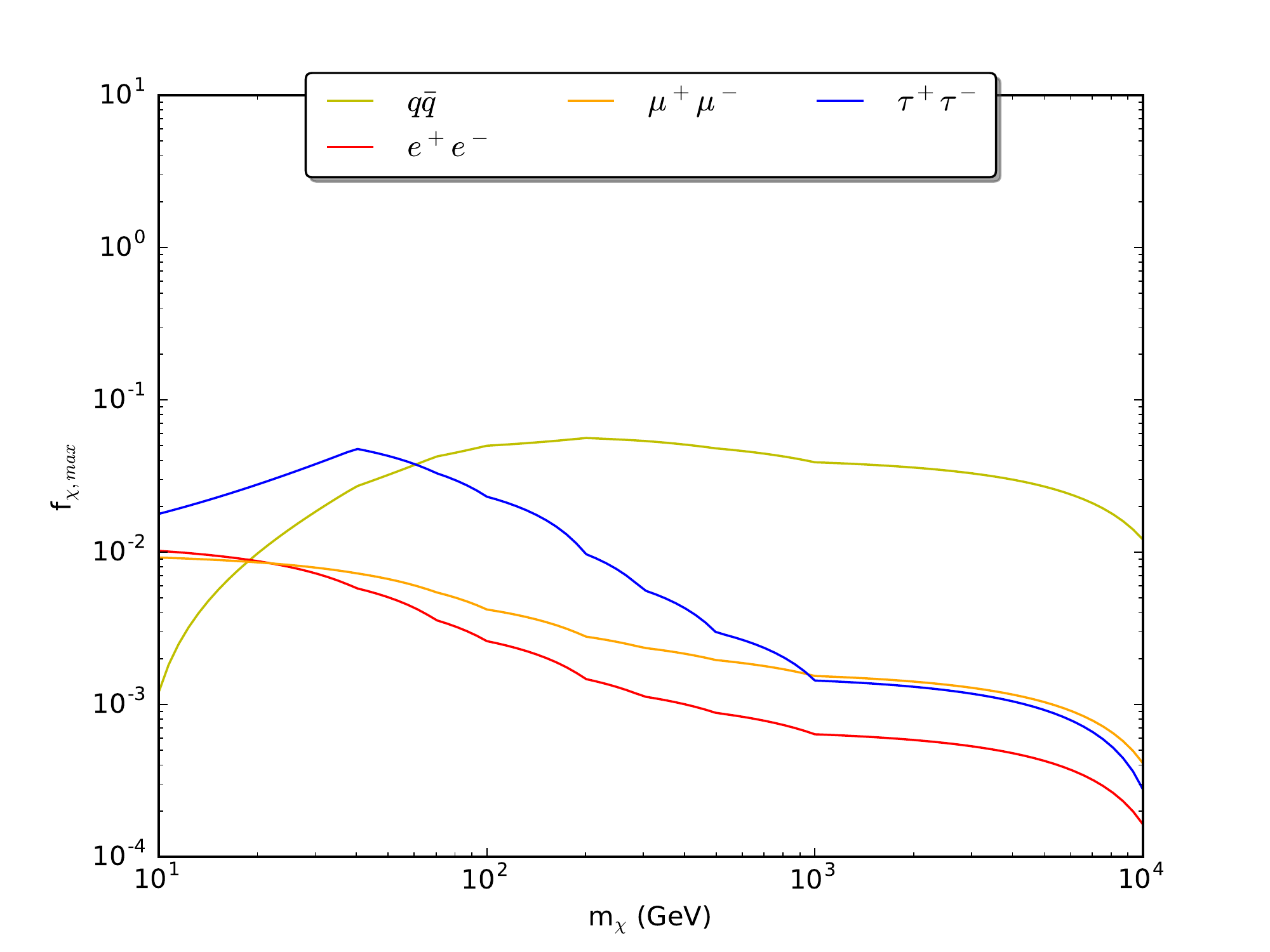}
	\includegraphics[width=0.45\textwidth]{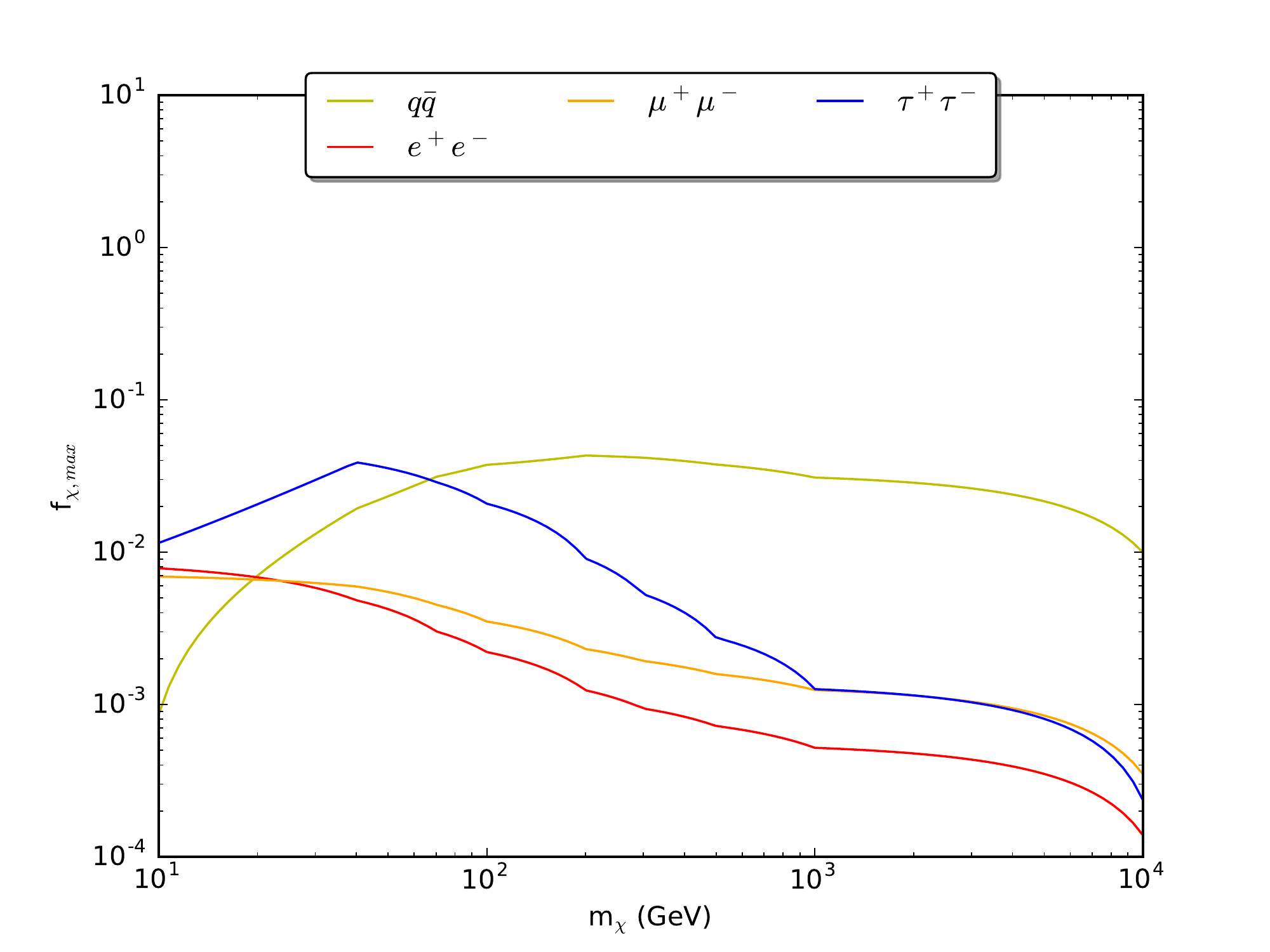}
	\caption{Maximum fraction of integrated gamma-ray flux accounted for by DM within M31 and Galactic Centre targets with contracted NFW halos for fermionic channels. Limits are found by imposing cross-section constraints from Coma diffuse radio data~\cite{coma-radio2003}. Left: M31 galaxy. Right: Galactic Centre.}
	\label{fig:radio-conf}
\end{figure}

\begin{figure}[ht!]
	\centering
	\includegraphics[width=0.45\textwidth]{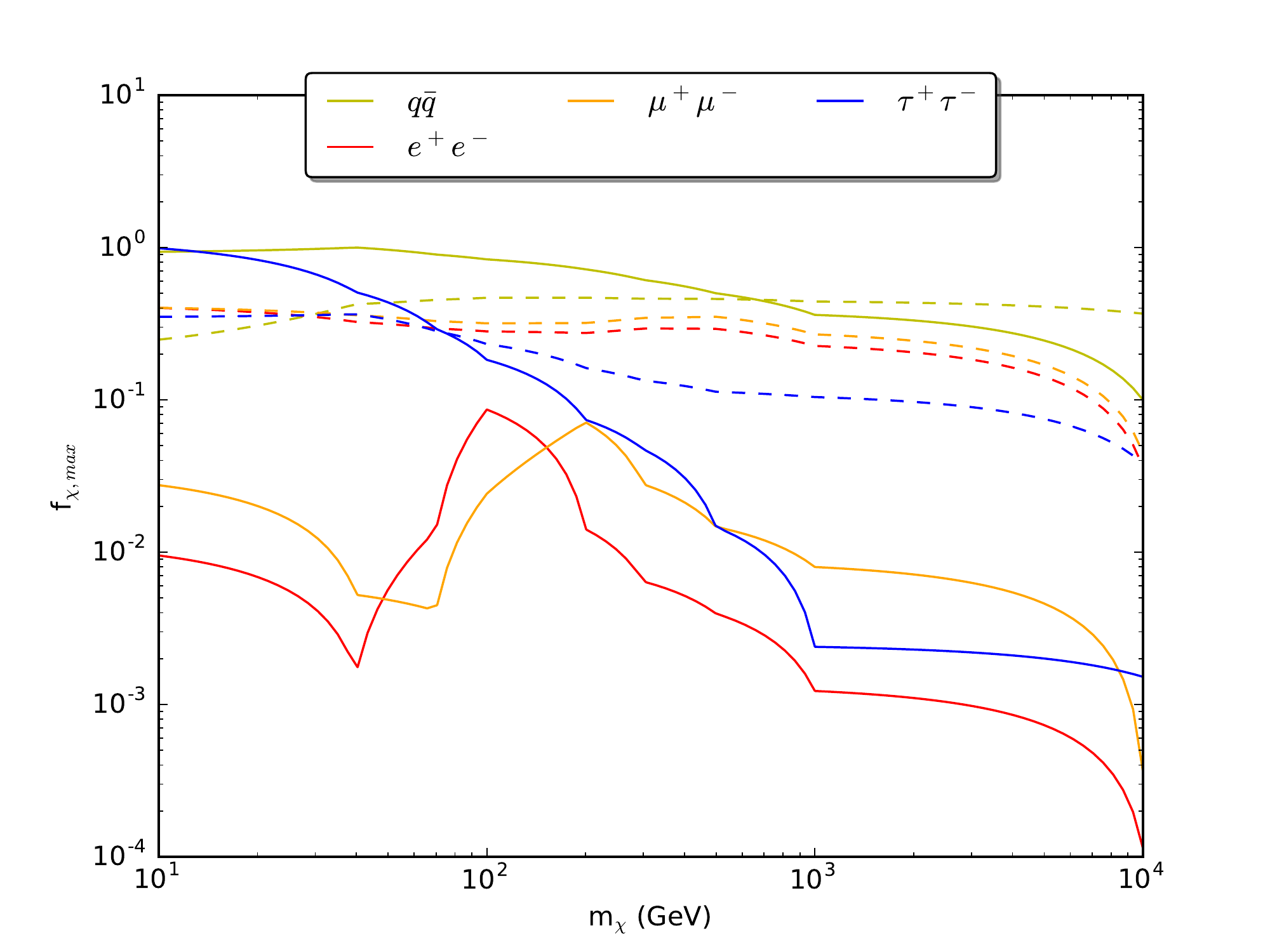}
	\includegraphics[width=0.45\textwidth]{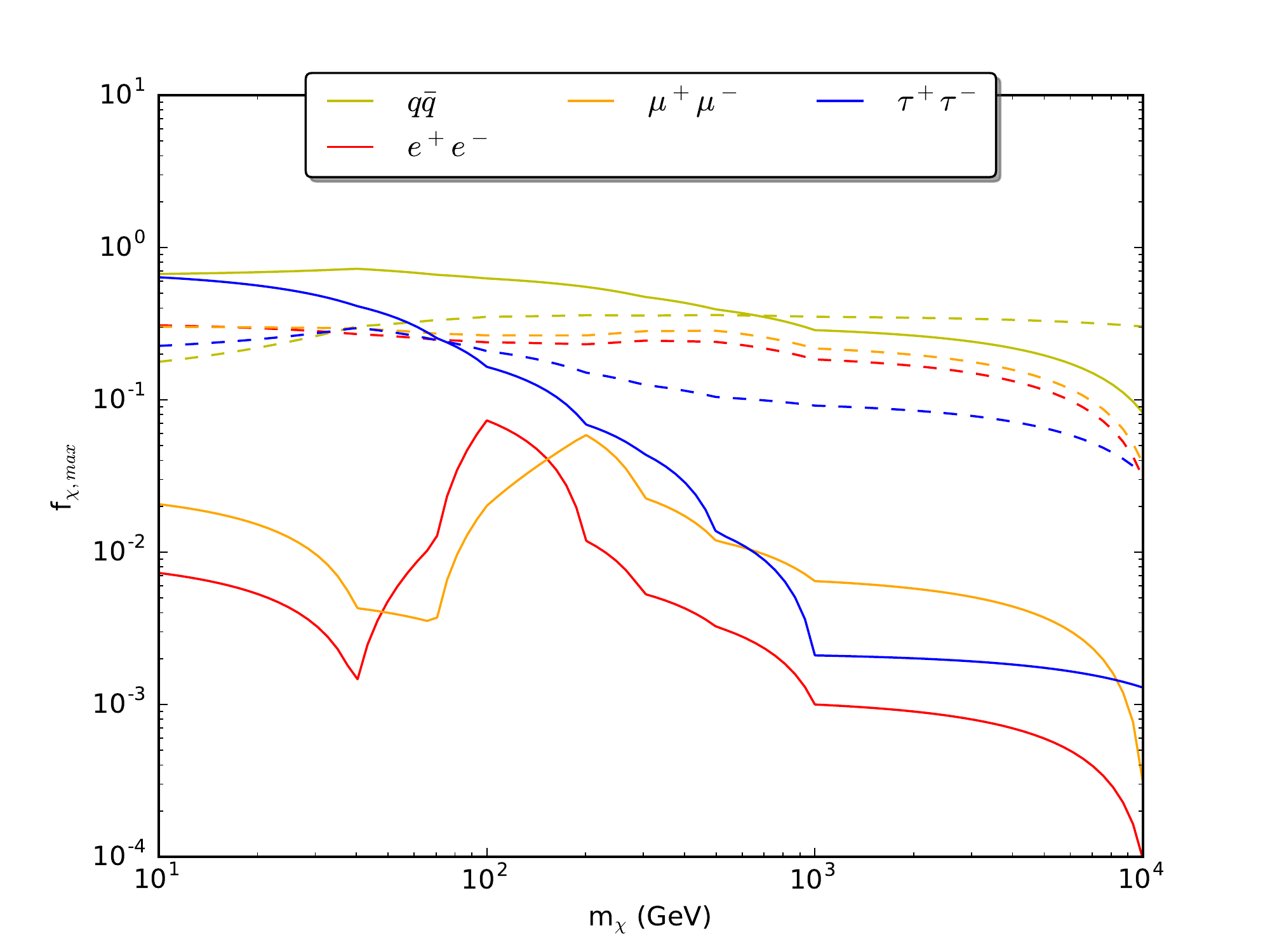}
	\caption{Maximum fraction of integrated gamma-ray flux accounted for by DM within M31 and Galactic Centre targets with contracted NFW halos for fermionic channels. Limits are found by imposing cross-section constraints from Fermi-LAT data on Coma~\cite{fermicoma2015} (solid lines) and Reticulum II~\cite{Fermidwarves2015} (dashed lines). Left: M31 galaxy. Right: Galactic Centre.}
	\label{fig:gamma-conf}
\end{figure}

In contrast with the preceding results, Figures~\ref{fig:radio-conb}, \ref{fig:radio-conf}, \ref{fig:gamma-conb}, and \ref{fig:gamma-conf} show the results analogous to Figs.~\ref{fig:radiob} to \ref{fig:gammaf}. However, this is when a contracted NFW profile from Eq.~(\ref{eq:nfwcon}) is assumed for both the GC and M31. Again, we show the maximal gamma-ray fraction accounted for by DM in M31 (left panels) and the Galactic Centre (right panels). In the case of the radio data in Figs.~\ref{fig:radio-conb} and \ref{fig:radio-conf}, we see that the imposed cross-sections from diffuse radio emission in Coma data are still strong, limiting all channels to a $\lesssim \mathcal{O}(10\%)$ contribution over the whole range of WIMP masses and for both targets. However, the limits from gamma-ray data are far more permissive, allowing nearly all of the gamma-ray spectrum to be accounted for by DM within the given energy range (apart from annihilation via low-mass leptons and the direct photon channel above $m_{\chi} = 30$ GeV).

\section{Discussion \& Conclusions}
\label{sec:conc}

The results presented here indicate the major factor in how much of the gamma-ray emissions within the GC and M31 can be accounted for by DM is the profile of the galactic halos. Despite this, the Coma diffuse radio data with the model of Coma employed suggest that, for any halo profile, DM is still relegated to a minor role in galactic gamma-ray emissions that in the studied energy range. However, given that some uncertainties must be introduced by the modeling assumptions (particularly halo substructure and magnetic field geometry), we supplemented the Coma results with those from the Reticulum II dwarf galaxy. As, in the case of this dwarf galaxy, there are no substructural effects accounted for here, and uncertainties are contributed largely by the $J$-factor. The constraints from gamma-rays in the Reticulum II environment were more permissive than the Coma radio data but still manage to substantially limit the contribution to galactic gamma-ray emission in M31 and GC when NFW or shallower halo profiles are assumed, imposing a limit of a DM contribution to the levels of $\mathcal{O}$(1\%). This held in both the cases where total integrated or bin-by-bin fluxes where used. Importantly, the results presented here using constraints from Coma and Reticulum II are quite consistent, despite the use of multi-frequency data subject to two independent sets of uncertainties (for radio and gamma-ray spectra respectively). This demonstrates the robustness of limits derived in this manner. An important question is the magnitude of the uncertainties that are propagated through into the maximal DM contribution fraction from various aspects of halo modelling for the Coma cluster and Reticulum II dwarf, where our cross-section limits are derived. In the case of the Coma cluster we take these uncertainties to be sourced from the mass of the cluster, the magnetic field strength, and the amplitude of the substructure boosting effect. Thus, we find that the DM galactic gamma-ray flux contribution limits derived from Coma have a relative uncertainty factor of $\sim 2$. While, in Reticulum II, we consider only the $J$-factor as source of uncertainty and find a relative uncertainty factor of $\sim 1.5$. These uncertainties are clearly not sufficient to mitigate the strength of the results for halo profiles in the GC and M31 that are NFW or shallower. 

When a contracted NFW halo is employed for M31 and GC, both the Coma and Reticulum II gamma-ray upper-limits are not sufficient to rule out a wide range of annihilation channels from making substantial $10 - 100\%$ contributions to the galactic gamma-ray spectra. It must be noted that the gamma-ray data points used are merely upper-limits, whereas the radio data being used constitutes measured diffuse fluxes (with the necessary caveat of possibly unresolved point sources). This leaves open the possibility that the modeling uncertainties in the case of Coma radio emission, which might allow for the limits from Figures like ~\ref{fig:radio-conb} and \ref{fig:gamma-conb} to be reconciled, could be obviated by an improvement in future gamma-ray limits on both Coma and Reticulum II reducing the permissiveness of the maximal DM contribution in M31 and the GC.

Another aspect to consider is whether the contracted NFW profile is a reasonable choice for the Milky-Way or Andromeda. Multiple results in the literature have suggested that gas cooling could result in the contraction of a DM halo~\cite{blumenthal1986,ryden1987,gnedin2004,gnedin2011}. However, the effects of bars may counter-act this~\cite{weinberg2002,weinberg2007} and it has been shown that an isothermal profile is favoured for M31~\cite{banerjee}, while a Burkert halo~\cite{burkert1995} may better fit the virial mass of the Milky-Way than a cusped profile~\cite{nesti2013}. This is supplemented by evidence that cored halo profiles are generally better at fitting data from structures smaller than galaxy clusters, like dwarf galaxies and low-surface brightness galaxies~\cite{salucci2001,walker2009,adams2014}, although this has been shown to hold to much larger galaxies~\cite{rodrigues2017}. In addition to this, it has been shown that cored galactic halo profiles may naturally evolve from cusps~\cite{governato2010,maccio2012}, powered by supernovae-generated gas out-flows~\cite{governato2012}. Thus, if the halos of the GC and M31 prove not to be steeply contracted cusps then, in both environments, even the NFW results shown here are enough to largely eliminate DM as more than a bit player in the emission of high-energy gamma-rays within these environments given current data on Coma and Reticulum II. This combines with recent work~\cite{ploeg2017} to strengthen the case for milli-second pulsars as an explanation of Galactic Centre gamma-ray emission above expected background levels. 

\section*{Acknowledgments}
We thank the Referee for several useful comments and suggestions.
SC acknowledges support by the South African Research Chairs Initiative
of the Department of Science and Technology and National
Research Foundation, as well as the Square Kilometre Array (SKA).
This work is based on the research supported by the South African
Research Chairs Initiative of the Department of Science and Technology
and National Research Foundation of South Africa (Grant
No 77948). GB acknowledges support from the DST/NRF
SKA post-graduate bursary initiative.

\end{document}